\documentclass[aps,prl,twocolumn,superscriptaddress,showpacs]{revtex4}

\usepackage{amsmath}
\usepackage{amssymb}
\usepackage{bm} 
\usepackage{epsfig}
\usepackage{graphicx}
\newcommand{\be}{\begin{equation}}
\newcommand{\ee}{\end{equation}}
\begin{document}

\title{Conformal Invariance and SLE  in Two-Dimensional Ising Spin Glasses }
\author{C.\ Amoruso}
\affiliation{School of Physics and Astronomy, University of Manchester,
Manchester M13 9PL, U.\ K.}
\author{A.\ K.\ Hartmann}
\affiliation{Institut f\"{u}r Theoretische Physik,
 Universit\"{a}t G\"{o}ttingen, Friedrich-Hund-Platz 1, 
37077, G\"{o}ttingen, Germany}
\author{M.\ B.\ Hastings}
\affiliation{Center for Nonlinear Studies and Theoretical Division, Los Alamos
National Laboratory, Los Alamos, NM 87545}
\author{M.\ A.\ Moore}
\affiliation{School of Physics and Astronomy, University of Manchester,
Manchester M13 9PL, U.\ K.}

\date{\today}

\begin{abstract}
We  present numerical evidence  that  the techniques  of conformal
field theory might be applicable to two-dimensional Ising spin glasses
with Gaussian bond distributions.   
It is shown that   certain domain wall  distributions
 in one  geometry can be
related to that in a second geometry by a conformal transformation. We
also present direct evidence  that the  domain walls are  
 stochastic
Loewner (SLE) processes with $\kappa \approx 2.1$. An argument is given that
their  
fractal dimension $d_f$ is related
to their interface energy exponent $\theta$ by
$d_f-1=3/[4(3+\theta)]$, which is  consistent with  the  commonly
quoted values $d_f \approx 1.27$ and $\theta \approx -0.28$.

\end{abstract}


\maketitle

The  powerful tools  provided  by conformal  field  theory (CFT)  have
 permitted  the determination  of the  exponents associated  with most
 two-dimensional (2d) critical  phenomena  \cite{CFT}. 
Unfortunately, CFT has
 not to date provided any results on systems like spin
 glasses, which  in two dimensions have  a zero temperature transition
 i.e. $T_c=0$. 
 As the temperature  of the system $T$ is reduced to
 zero, the  correlation length $\xi$ increases  to infinity \cite{KLY,
 droplet} as $\xi(T) \sim 1/ T^{\nu}$.  
 The   Hamiltonian    of   the   system    is   
$ H=-\sum_{<ij>}J_{ij}S_iS_j$,
with  $S_i=\pm 1$.  If the nearest-neighbour
bond   distribution  $J_{ij}$   is  continuous,   as  in the  Gaussian
distribution, then   the exponent  $\nu=-1/\theta$, where  $\theta$  is the
exponent which  describes how the energy  $\Delta E$ of  a domain wall
(DW)
which crosses  a system of linear  extent $L$ depends upon  $L$: $\Delta E
\sim  L^{\theta}$ \cite{droplet}.   According   to  numerical  studies 
 \cite{Carter,
Hartmann} $\theta$ lies between -0.28 and -0.29. The DW has a
fractal dimension $d_f \approx 1.27  \pm 0.01$ \cite{HY}. 
  While the  disorder present in spin glasses 
means that at a local level there is not even translational 
invariance, the existence of a diverging lengthscale such as $\xi(T)$
suggests that on long lengthscales such microscopic features 
could become irrelevant and that possibly even CI might arise.
  It is the chief purpose here to provide numerical evidence
 that this is indeed the case and so hopefully pave the way to eventually
 determining
exponents like $\theta$ using CFT.

 In the first part of this letter we present
a  numerical  study  of  whether there  is  conformal
invariance  of the  DW  distribution in  2d spin
glasses. Within our numerical accuracy, CI does seem
to hold  in the thermodynamic limit.  Then we next
present numerical evidence  that DWs in 2d
are stochastic Loewner evolution
(SLE)  processes  \cite{Cardy},  and finally  we suggest a relationship 
 between $\theta$ and  $d_f$.

CI of  the  DW  distribution implies  that
given two  geometries related by a conformal  transformation, then the
probability of finding the DW in a given configuration in one
geometry  is related  to the  probability of  finding  the conformally
transformed DW configuration in the transformed geometry.  We
will find a transformation $F(z)$ mapping geometry $(a)$ onto $(d)$ in
Fig.\ \ref{fig:mapping}.  
The two rectangles  have periodic boundary conditions so that
the  left and right  edges are  identified, while  the top  and bottom
edges  are open i.e. they have the topology of an annulus.   The  two slits
  in  $(d)$ also  have open  boundary
conditions, so that no bonds cross the slits.  The rectangle $(a)$ has
an arbitrary aspect ratio; we  can tune the distance between the slits
in  $(d)$ as desired,  with the  aspect ratio  of the  right rectangle
being  a function of  the distance  between the  slits and  the aspect
ratio    of   the    left   rectangle    as   given    implicitly   by
Eq.~(\ref{implicit}).  The  dashed and  dotted lines in  each geometry
are also conformally mapped to each other.


Before presenting the desired conformal mapping,
we discuss the implications of CI in these
geometries.
First,  we  present   the  implications   of  conformal
invariance  for the  probability distribution  of the  domain  wall in
these two geometries.  We can  measure the probability $p_1(n)$ of the
domain  wall crossing  the  dashed  line $n$  times  in the  rectangle
geometry, $(a)$,  as well as  the probability $p_1'(n)$ of  the domain
wall  crossing the line  $n$ times  in the  slit geometry,  $(d)$.  In
Fig.\ \ref{fig:mapping}, 
we also show the mapping of  a DW which crosses the dashed
line three  times.  
In addition,  we can  measure the probability  $p_2(n)$ of  the domain
wall crossing the dotted line  $n$ times in the rectangle geometry, as
well  as  the  probability  $p_2'(n)$  of $n$  crossings  in  the  slit
geometry.  By  statistical translational invariance,  this probability
is equal  to $p_1(n)$.
Thus, a naive application  of CI
would suggest that $p_2'(n)=p_1'(n)$.   However, in the continuum limit
it is not  possible to distinguish between $n$  and $n+2$ crossings of
the  dashed line  for $n\neq  0$; if  the distance  between successive
crossings of the dashed line in  Fig.\ \ref{fig:mapping} 
is of the order of the lattice
spacing,  this looks  at large  scale like  a single  crossing  of the
dashed line.  Thus, the predictions of CI are:

\begin{figure}[tb]
\centerline{ \includegraphics[width=0.95\columnwidth]{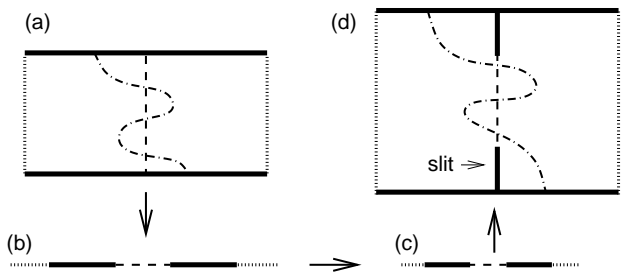} }
\caption{Mapping between different geometries, and 
of the DW, shown as the dashed-dotted line.}
\label{fig:mapping}
\end{figure}

\begin{eqnarray}
\label{conf2}
p_1^{\prime} \equiv \sum_{n \, \rm even}  p_1'(n) =  \sum_{n \,  \rm even}
p_2'(n)\equiv p_2^{\prime},  \nonumber \\ 
\sum_{n \, \rm  odd}  p_1'(n)=\sum_{n \,  \rm odd}  p_2'(n), \quad p_1'(0)=p_2'(0).
\end{eqnarray}
These  identities in Eq. (\ref{conf2})  are  the  consequence  of  a  conformal
automorphism of  the slit geometry  which interchanges the  dotted and
dashed lines: $z\rightarrow F(F^{-1}(z)+\pi)$. 
We explicitly checked for   a small rectangular system 
($N=30*30$)  that $p1 =p2 \sim 0.603$ 
and they are equal to  $p_2^{\prime}$ (quoted in tables 1 and 2)
within the statistical error.

Our main test for CI is to find whether $p_1'=p_2'$.
The naive
expectation is that the DW is less likely to cross
the dashed line,
due to the constriction reducing the number of possible configurations,
while CI instead requires these probabilities to be equal.



{\it  Constructing  the  Conformal  Transformation---} We  define  the
rectangle $(a)$ to  have width $2\pi$ and height  $h$, and thus aspect
ratio  $h/2\pi$.  The  horizontal  direction as  plotted  is the  real
coordinate  running from  $-\pi$ to  $\pi$ while  the vertical  is the
imaginary coordinate,  which runs from  $-ih/2$ to $ih/2$.   We define
$\tau(h)=2\pi  i/h$,  and $\lambda(\tau)$  to  be  the modular  lambda
elliptic function. 
   We define a function $g(z)$  to map from $(a)$
to    $(b)$ in    Fig.\ \ref{fig:mapping} 
   by    \be    g_h(z)={\rm     sn}(2    i    z
K(\lambda(\tau(h)))/h|\lambda(\tau(h))), \ee where $K$ is the complete
elliptic integral  of the first kind  and $sn$ is  the Jacobi elliptic
function \cite{kober}.   This function $g$ maps the  dashed and dotted
lines as shown.  The dotted lines extend off to infinity and all lines
lie on the real axis.  The  upper and lower lines of the rectangle are
mapped to  the two solid  lines, which have  endpoints at $\pm  1$ and
$\pm 1/\sqrt{\lambda(\tau(h))}$.

The mapping from $(b)$ to $(c)$ is simply $z\rightarrow s z$, for some
parameter $0<s\leq 1$; smaller values of $s$ produces deeper cuts into
the rectangle in  $(d)$.  In $(c)$, the endpoints of  the lines are at
$\pm  s$  and $\pm  s/\sqrt{\lambda(\tau)}$.   We  then determine  the
height $h'$ of the rectangle in $(d)$ such that
 \be
\label{implicit}
1/\sqrt{\lambda(\tau(h'))}=s/\sqrt{\lambda(\tau(h))}.   \ee 
 The final
mapping from  $(c)$ to  $(d)$ is $z\rightarrow  g_{h'}^{-1}(z)$.  This
maps the portion of the solid  lines in $(c)$ between $\pm 1$ and $\pm
1/\sqrt{\lambda(\tau(h'))}$ onto  the upper  and lower borders  of the
rectangle in $(d)$, while the portion  of the solid lines in the third
geometry between $\pm s$ and $\pm 1$ are mapped onto the slits.  Thus,
the  full  mapping  from  $(a)$  to $(d)$  is  \be  F(z)=g_{h'}^{-1}(s
g_h(z)), \ee  and the endpoints of  the slits in $(d)$  are located at
$F(\pm ih/2)$.


We proceed  by first  finding the ground state  of the system,  using a
mapping  to  a graph-theoretical  problem,  the minimum-weight  perfect
matching problem \cite{HR}. Domain  walls were created by flipping the
signs of the  horizontal bonds in a column.  Because of the periodicity
in the  horizontal direction,  this induces a  DW  across the
system and  it is  the crossings  of the dashed  central line  and the
dotted \lq end' line which we  study. A DW is best defined as
a walk on the lattice dual to the original lattice, and the dotted and
dashed lines are also lines of this dual lattice.

In Tables 1 and 2 are displayed the values of $p^{\prime}_1$ and
$p^{\prime}_2$ and the probabilities of zero crossings, $p_1'(0)$ and
$p_2'(0)$ for $s=0.95$ and $s=0.90$ for various \lq sizes'. Thus
$30*32$ means that the system studied is rectangular with 30 spins on
each horizontal line (the direction in which the system is periodic)
and 32 spins on each vertical line. The next number 8 is the number of
rows cut by slits (4 at the top of the system, 4 at the bottom of the
system) indicating that there are $(32 - 8)$ rows not cut by slits.
(For small sizes it is not possible to find integers to get
the sizes precisely correct for the given the aspect ratio). 

\begin{table}[ht]
TABLE 1: Approach to CI for $s=0.95$
\begin{tabular}{l|rrrrr}
SIZE & $p_1'$   &$p_2'$ & $p_1'(0)$& $p_2'(0)$ & samples      \\\hline
$30*32;\ 8  $   & 0.624    &  0.603 & 0.463  & 0.409  & 20000 \\
$52*55;\ 14 $   & 0.618    &  0.609 & 0.424  & 0.392  & 10000 \\
$76*80;\ 20 $   & 0.616    &  0.615 & 0.417  & 0.385  & 10000 \\
$98*103;\ 26$   & 0.617    &  0.615 & 0.414  & 0.385  &  7000 \\
\end{tabular}
\end{table}
\begin{table}[ht]
TABLE 2: Approach to CI for $s=0.90$
\begin{tabular}{l|rrrrr}
SIZE & $p_1'$   &$p_2'$ & $p_1'(0)$& $p_2'(0)$ & samples            \\\hline
$30*33;\ 12$    &  0.639   & 0.604   & 0.494  & 0.407  &     20000  \\
$50*56;\ 20$    &  0.623   & 0.612   & 0.448  & 0.399  &     10000  \\
$98*109;\ 38$   &  0.620   & 0.611   & 0.421  & 0.384  &      6000  \\
$124*138;\ 48$  &  0.612   & 0.610   & 0.411  & 0.381  &      6000  \\
\end{tabular}
\end{table}

\begin{figure}[tb]
\centerline{ \includegraphics[width=0.8\columnwidth]{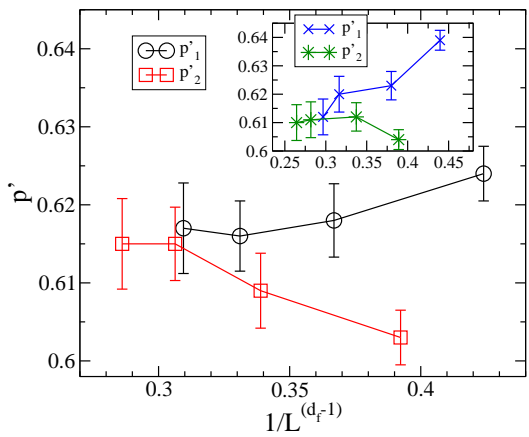} }
\caption{Approach of $p_1'$ and $p_2'$ to asymptopia. Main figure is
  for $s=0.95$ while inset for $s=0.90$.} 
\label{fig:p1p2}
\end{figure}

For each probability $p$ in these tables  $\sqrt(p(1-p)/N_s)$ is its
standard deviation,
where $N_s$ is the number of samples i.e. bond realizations averaged
over.  As the size increases so the continuum limit is
approached, the closer  $p_1'$ and $p_2'$ become, implying that the
distribution of the DWs is conformal. In Fig.\ \ref{fig:p1p2} we have
plotted $p_1'$ versus $1/L_1^{d_f-1}$ and $p_2'$ versus
$1/L_2^{d_f-1}$. (We have no proof that this is the way that $p_1'$
and $p_2'$ approach their asymptotic limit, but this dependence will
be partly motivated below). Again one can see that in the continuum
limit CI seems to hold. $L_1$ is the number of rows
not cut by the slit and $L_2$ is the total number of rows in the
system. Thus for the 30*32 system size, $L_1=24$ and $L_2=32$.

\begin{figure}[tb]
\centerline{ \includegraphics[width= 0.8\columnwidth]{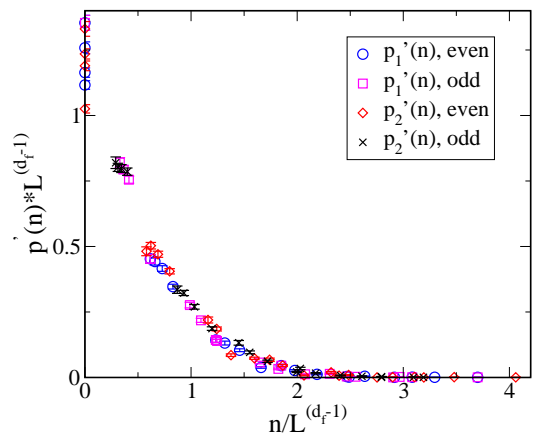} }
\caption{Scaling of probability of crossings for $s=0.95$. }
\label{fig:scaling}
\end{figure}

\begin{figure}[tb]
\centerline{ \includegraphics[width= \columnwidth]{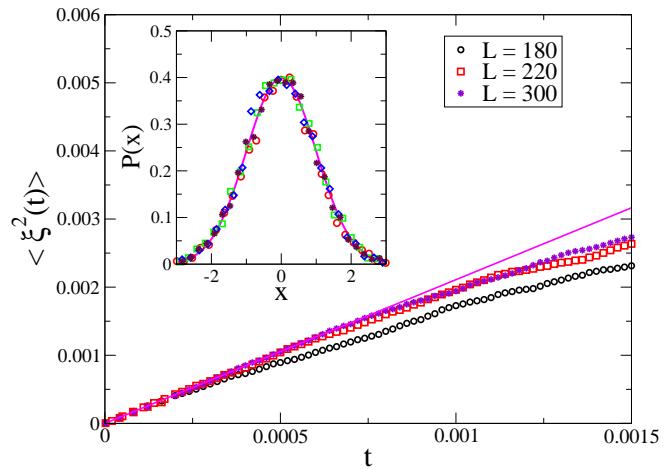} }
\caption{$\langle \xi(t)^2\rangle$ versus Loewner time $t$ for three different
system sizes $L$. The straight line has slope $2.1$, which is our estimate for
$\kappa$. Inset: probability distribution at four different times,
$ 0.0003,0.0004, 0.0005$ and $0.0006$  when
$L=300$. The curve
is the Gaussian $\exp(-x^2)/\sqrt{2\pi}$, where $x=\xi(t)/\sqrt{\kappa t}$. }
\label{fig:SLE}
\end{figure}

 Unfortunately the probabilities for zero crossings $p_1'(0)$ and
$p_2'(0)$ appear to approach each other very slowly (if at all). In an
attempt to understand this behaviour, we have studied the probabilities
of $n$ crossings, $p_1'(n)$, of the central (dashed) line, and
$p_2'(n)$ of the right (dotted) line.  $p_1'(n)$  has a scaling
dependence on the number of crossings $n$ as
$1/L_1^{d_f-1}f_a(n/L_1^{d_f-1})$ and similarly $p_2'(n)$ is of the
form $1/L_2^{d_f-1}f_a(n/L_2^{d_f-1})$. The subscript $a$ is added (so
$a=e$ or $a=o$) to allow us to distinguish even values of $n$ from odd
values, as the DW is topologically very different depending
on the parity of $n$.  $ p_1'(n\neq 0)$ can mean that macroscopically
the DW crosses the central line once (say), but then if one
zooms in on that single crossing, it actually crosses many times.  To
see how many times it would cross, suppose the DW has fractal
dimension $d_f$.  Then, the intersection of the DW and a
vertical line has dimension $d_f-1$.  Thus, the expected number of
crossings would be $L_1^{d_f-1}$ which gives the above scaling forms.
In Fig.\ \ref{fig:scaling} we display the data.
The scaling functions for even and
odd seem to be very similar, even though the DWs are
topologically very different.  Notice that the no-crossing
probabilities are clearly not part of this scaling form (which they
cannot be if they are non-zero). However, we suspect in the light of
the above that the convergence of $p_1'(0)$ and $p_2'(0)$ to each
other might be as slow as $1/L^{d_f-1}$, and as $d_f$ is about 1.27
this could be a very slow convergence rate. We have also studied the
case of $s=0.85$, which corresponds to a still deeper cut by the
slits. For this case, the convergence of even $p_1'$ to $p_2'$ has not
been achieved in the largest sizes we have studied $(146*175)$, and we
suspect that the slow convergence here is of similar origin.
The conformal invariance found in the DW distribution encouraged us to find out
 if the domain walls were also SLE processes \cite{Cardy}. Suppose the domain
wall is the curve $\gamma(t)$ which begins at a point on the boundary of
 the upper half plane $H$. The half-plane $H$ minus the curve $\gamma(t)$ can
be mapped back onto $H$ by an analytic function $g_{t}(z)$ which is made 
unique by demanding that $g_t(z) \sim z+2t/z+\rm{O}(1/z^2)$ at infinity.
 The growing tip of the curve is mapped onto the real point $\xi(t)$. The
 DW is an SLE process if $\xi(t)$ is a Brownian walk whose elements have an
 independent Gaussian distribution and $\langle \xi(t)^2\rangle=
\kappa t$. The diffusion coefficient $\kappa$ is of prime importance as it is
 related to the central charge of the conformal field theory \cite{Cardy}. 
 In practice we approximate $g_t(z)$ by composing a sequence
 of discrete, conformal slit maps of the
form $z\rightarrow \sqrt{(z-\xi_i)^2+4\Delta t_i}+\xi_i$,
where the parameters $\Delta t_i,\zeta_i$ are chosen so that
the $i$-th such map removes the $i$-th step
from the domain wall following
 the procedures of Ref. \cite{Bernard}, to produce a series
 of times $t_i=t_{i-1}+\Delta t_i$ and values $\xi(t_i)=\xi_i$
which approximate $\xi(t)$.
We denote the coordinates of the domain wall points by $z_i^0$, $i=1...N$.
The first such slit map transforms the coordinate of the first
step, $z_1^0$, into the origin, and transforms
$z_j^0$, $j>1$, into a new point $z_j^1$.
In general, the $j$-th map transforms $z_j^{j-1}$ into the origin
and gives
\begin{eqnarray}
\label{eqn:map}
t_i &=& t_{i-1} + (y_i^{i-1})^2/4, \quad \xi(t_i)  =  x_i^{i-1}\\ 
z^{i}_{j}  &=&  \sqrt{[z^{i-1}_j - x_i^{i-1}]^2 + (y_i^{i-1})^2} + x_i^{i-1},
\, (j>i). \nonumber
\end{eqnarray}
As usual the complex number $z =(x,y)$ and $\xi(t_0)=0$.
 The sign of the square root is chosen
 so that it has the sign of $[x_j^{i-1}- x_i^{i-1}]$.
The geometry
which we studied was a square $L \times L$ with periodic boundary conditions
in the horizontal direction and open boundaries in the vertical direction
which is the direction in which the domain runs.  In Fig. \ref{fig:SLE} we
 show the average over  realizations of the
disorder of  $\langle \xi(t)^2\rangle$, plotted against t, for three values of
$L$. When $L=180$, we took 3000 disorder realizations, $L=220$, 4000 and
 for $L= 300$
we took 5000.  $\langle \xi(t)^2\rangle$ is linear in time for a range of
 times 
which increase with the
 system size $L$ and from the slope of this linear region we estimate 
that $\kappa \approx 2.1$. Our
boundary conditions, together with the finite size of the system, means that
 it does not properly satisfy the requirements for producing either chordal
 or dipolar SLE \cite{Cardy}, which may partly explain the modest size of
 the linear
 regime. In a recent related study it was found  that 
using the dipolar SLE did indeed extend the linear regime \cite{BDM}. In the 
inset to Fig. \ref{fig:SLE} we show that the probability distributions of 
$\xi(t)/\sqrt{\kappa t}$ at four different times within
the linear regime are  standard Gaussian  as would be required for
 the domain walls
 to be SLE processes. 

If spin-glass  domain walls are SLE processes, there may be 
  a relationship between the
fractal dimension of the DW $d_f$ and the exponent $\theta$.      The 
   fractal     dimension    is
related to $\kappa$ via
$d_f=1+\kappa/8$ \cite{Cardy}. (Our numerical value for $\kappa
 \approx 2.1$ 
 and the estimates of $d_f \approx 1.27\pm0.01$ in Ref. \cite{HY} are
 consistent with this 
relationship).
The correlation length exponent exponent $\nu$ is related to one of
the Kac elements of conformal field theory: 
for example in Potts models with components $Q$, $1\leq Q \leq 4$, one 
has $d-1/\nu = 2+\theta=2h_{2,1}$, but in   
other models $d-1/\nu$ is $2 h_{1,3}$ or $2 h_{1,2}$. $d$ is
 the dimensionality of the system i.e. 2.
 Now if SLE applies, each of these elements is related to $\kappa$:
$2 h_{2,1}=(6-\kappa)/\kappa$ \cite{Cardy}, $2h_{1,3}=\kappa-2$ etc.
 For each of these 
possibilities one can derive a  relationship between $d_f$ and
$\theta$  and the only one
which comes close to the numerical values $d_f=1.27 \pm 0.01$ 
and $\theta=-0.285 \pm 0.05$ is from
$2h_{2,1}=\frac{6-\kappa}{\kappa}=d-\frac{1}{\nu}=d+\theta$.
Then on  eliminating $\kappa$ in favour of $d_f$ gives 
\be 
d_f=1+\frac{3}{4(3+\theta)}.
\label{dstheta}
\ee
On using one of the alternative possible relations, say $2+\theta =2h_{1,3}$,
 $d_f=(12+\theta)/8$.
 Then the predicted
value of $d_f =1.46$ which is not consistent with its numerical value. 
 Eq. (\ref{dstheta}) seems to be  the only possible relationship between
 $d_f$ and $\theta$  which is  compatible with their
numerically well-established values. (Note that Eq. \ref{dstheta} would not
 apply
to the $\pm$J spin-glass  model as for it $\nu$ might have the same
 value as for  the spin glass model whose bonds have a Gaussian
 distribution \cite{Marinari}
 but it has $\theta=0$ \cite{YH}). The apparent success of Eq. \ref{dstheta}
in providing a  relationship between $d_f$ and $\theta$ might provide a
 clue in finding the kind of 
conformal field theory appropriate for two-dimensional spin glasses. Our
numerical evidence for conformal invariance and SLE strongly suggests that
 such a field theory should indeed exist.

\begin{acknowledgments}
We would like to thank Antonio Celani,   
Andreas Honecker, Ron Fisch and Marco Picco for fruitful discussions. 
We should like to thank the authors of Ref. \cite{BDM} for 
communicating the details of their numerous SLE investigations.
This work was supported from the European Community via the DYGLAGEMEM
contract.
AKH acknowledges financial support from the
{\em VolkswagenStiftung} (Germany) within the program
``Nachwuchsgruppen an Universit\"aten''.  MBH was supported by 
DOE W-7405-ENG-36.
\end{acknowledgments}


\begin{thebibliography}{99}
\bibitem{CFT} P. Di Francesco, P. Mathieu and D. Senechal
  \textit{Conformal field theory} 
(Springer, 1997).



\bibitem{KLY} H. G. Katzgraber, L. W. Lee and A. P. Young,
 Phys. Rev. B \textbf{70}, 014417 (2004).

\bibitem{droplet} A.  J.  Bray  and M.  A.  Moore, in  \textit{ Glassy
Dynamics  and  Optimization},  edited   by  J.   L.   van  Hemmen  and
I. Morgenstern, (Springer, Berlin, 1986); D. S. Fisher and D. A. Huse,
Phys. Rev.   Lett. \textbf{ 56},  1601 (1986); Phys. Rev.   B \textbf{
38}, 386 (1988); W. L. McMillan, J. Phys. C \textbf{ 17}, 3179 (1984).

\bibitem{Carter} A. C. Carter, A. J. Bray and M. A. Moore, Phys. Rev. Lett.
\textbf{88}, 077201 (2002).

\bibitem{Hartmann} A. K. Hartmann, A. J. Bray, A. C. Carter, M. A. Moore and
A. P. Young, Phys. Rev. B \textbf{66}, 224401 (2002).

\bibitem{HY} A. K. Hartmann and A. P. Young, Phys. Rev. B \textbf{66}, 094419
 (2002).

\bibitem{Cardy} For a reviews aimed at physicists,
 see J. Cardy, Annals of Physics, \textbf{318}, 81 (2005); M. Bauer
 and D. Bernard, preprint math-ph/0602049.

\bibitem{kober} Function $g_h(z)$ is related by a logarithmic map to one given
in
H. Kober, \textit {Dictionary of Conformal Representations} 
(Dover 1957), section 13.11, p. 192.

\bibitem {HR} A. K. Hartmann and H. Rieger, {\em Optimization Algorithms in
Physics}, (Wiley-VCH, Berlin 2001).

\bibitem{Bernard} D. Bernard, G. Boffetta, A. Celani and G. Falkovich, 
Nature Physics \textbf{2}, 124 (2006).

\bibitem{BDM} D. Bernard, P. Le Doussal and A. A. Middleton, to be published.

\bibitem{Marinari} T. J\"{o}rg, J. Lukic, E. Marinari and O. C. Martin, 
Phys. Rev. Lett. \textbf{96}, 237205 (2006).

\bibitem{YH} A. P. Young and A. K. Hartmann, Phys. Rev. B \textbf{64}, 180404
 (2001).


\end{thebibliography}
\end{document}